\documentclass[aps,pra,twocolumn,superscriptaddress,nofootinbib]{revtex4-1}
\usepackage{amsmath}
\usepackage{amssymb}
\usepackage{graphicx}
\usepackage{mathrsfs}
\usepackage{ntheorem}
\usepackage{mathtools}
\usepackage{enumerate}
\usepackage{enumitem}
\usepackage{times,txfonts}
\usepackage{subfigure}
\usepackage{upgreek}
\usepackage{bm}
\usepackage{tikz}
\usepackage[colorlinks,bookmarksopen,bookmarksnumbered,citecolor=blue, linkcolor=blue, urlcolor=blue]{hyperref}
\newcommand{\ket}[1]{|#1\rangle}
\newcommand{\bra}[1]{\langle #1|}
\newcommand{\Tr}{\mathrm{Tr}}

\newcommand{\abs}[1]{\lvert #1\rvert}

\def\CC{{\rm\kern.24em \vrule width.04em height1.46ex depth-.07ex \kern-.30em C}}
\def\RR{{\rm\kern.24em \vrule width.04em height1.46ex depth-.07ex
\kern-.30em R}}
\def\P{{\rm I\kern-.25em P}}

\begin{document}

\title{Coherence filtration under strictly incoherent operations}

\author{C. L. Liu}
\email{clliusdu@foxmail.com}
\affiliation{Graduate School of China Academy of Engineering Physics, Beijing 100193, China}
\author{C. P. Sun}
\email{suncp@gscaep.ac.cn}
\affiliation{Graduate School of China Academy of Engineering Physics, Beijing 100193, China}
\affiliation{Beijing Computational Science Research Center, Beijing 100193, China}

\date{\today}
\begin{abstract}
We study the task of coherence filtration under strictly incoherent operations. The aim of this task is to transform a given state $\rho$ into another one $\rho^\prime$ whose fidelity with the maximally coherent state is maximal by using stochastic strictly incoherent operations. We find that the maximal fidelity between $\rho^\prime$ and the maximally coherent state is given by a multiple of the $\Delta$ robustness of coherence $R(\rho\|\Delta\rho):=\min\{\uplambda|\rho\leq\uplambda\Delta\rho\}$, which provides $R(\rho\|\Delta\rho)$ an operational interpretation. Finally, we provide a coherence measure based on the task of coherence filtration.
\end{abstract}
\maketitle

\section{Introduction}
Quantum coherence is among the necessary features of quantum mechanics for the departure between the classical and quantum world. It is an essential component in quantum information processing \cite{Nielsen}, and plays a central role in various fields, such as quantum computation \cite{Shor,Grover}, quantum cryptography \cite{Bennett},  quantum metrology \cite{Giovannetti,Giovannetti1}, and quantum biology \cite{Lambert}. Recently, the resource theory of coherence has attracted growing interest due to the rapid development of quantum information science \cite{Aberg1,Baumgratz,Streltsov,Fan,Wu}. The resource theory of coherence not only establishes a rigorous framework to quantify coherence, but also provides a platform to understand  quantum coherence from a different perspective.

Any quantum resource theory is characterized by two fundamental ingredients, namely, the free states and the free operations \cite{Horodecki,Brandao,Chitambar}. For the resource theory of coherence, the free states are quantum states which are diagonal in a prefixed reference basis. The free operations are not uniquely specified. Motivated by suitable practical considerations, several free operations have been presented \cite{Levi,Aberg1,Baumgratz,Vicente,Chitambar1,Chitambar2,Winter,Yadin,Marvian}. In this paper, we focus our attention on the strictly incoherent operations, which were first proposed in Ref. \cite{Winter} and, in Ref. \cite{Yadin}, it has been shown that these operations neither create nor use coherence and have a physical interpretation in terms of interferometry. Thus, the set of strictly incoherent operations is a physically well-motivated set of free operations for the resource theory of coherence.

In the resource theories, a remarkable effort has been devoted to studying various distillation protocols \cite{Chitambar}. The distillation process is the process that extracts pure resource states from a general state via free operations. For the resource theory of coherence, various coherence distillation protocols were proposed. These protocols can be divided into two classes: the asymptotic coherence distillation and the one-shot coherence distillation. The asymptotic coherence distillation of pure states and mixed states by using strictly incoherent operations was studied in Refs. \cite{Lami,Lami1,Zhao}. To relax several unreasonable assumptions of the asymptotic regime, i.e., unbounded number copies of identical states and collective operations, several one-shot coherence distillation protocols were proposed and explored \cite{Liu1,Liu2,Liu3,Fang,Regula,Regula1,Torun,Chen,Zhang,Bu,Liu4,Liu5}.

However, the literature just mentioned and the results therein suggested that the conditions of exact coherence distillation are too stringent. Inspired by the task of filtration of other quantum resource theories \cite{Chitambar,Dur}, which is a basic operational task in quantum resource theories \cite{Chitambar}, some authors started to consider the task of coherence filtration instead. Specifically, this task was studied under maximally incoherent operations and dephasing-covariant incoherent operations in Refs. \cite{Fang,Regula1}. The aim of coherence filtration is to transform a given state $\rho$ into another one $\rho^\prime$ whose fidelity with the maximally coherent state is maximal by using free operations. Although some related results about pure coherent states were obtained, the coherence filtration about mixed states under strictly incoherent operations has been relatively unexplored.

In this paper, we address this problem completely by developing the coherence filtration protocol under strictly incoherent operations. To this end, we first calculate the maximal fidelity between $\rho^\prime$ and the maximally coherent state and we find that it is given by a multiple of the $\Delta$ robustness of coherence, $R(\rho\|\Delta\rho)$, which was given in Ref. \cite{Chitambar2}. This provides an operational interpretation  to the $\Delta$ robustness of coherence. Finally, we obtain a coherence measure from this task and this further shows that the quantity $R(\rho\|\Delta\rho)$ can be viewed as a coherence monotone which extends the coherence rank to mixed states.

This paper is organized as follows. In Sec.~\ref{II}, we recall some notions of the resource theory of coherence. In Sec.~\ref{III}, we calculate the maximal fidelity between $\rho^\prime$ and the maximally coherent state. In Sec.~\ref{IV}, we present several results relating to the coherence filtration protocol. Section \ref{V} contains our conclusions.

\section{Preliminaries}\label{II}

To present our result clearly, it is instructive to introduce some elementary notions of the resource theory of coherence \cite{Baumgratz}.
Let $\{\ket{i}\}_{i=1}^d$ be the prefixed basis in the finite-dimensional Hilbert space. A state is said to be incoherent if it is diagonal in the basis and the set of such states is denoted by $\mathcal{I}$. Coherent states are states not of this form. For a pure state $\ket{\varphi}$, we will write $\varphi:=\ket{\varphi}\bra{\varphi}$. The $d$-dimensional maximally coherent state has the form
\begin{eqnarray}
\ket{\psi_d}=\frac1{\sqrt{d}}\sum_{i=1}^{d}\ket{i}.
\end{eqnarray}

A strictly incoherent operation \cite{Winter, Yadin} is a completely positive trace preserving map, expressed as
\begin{eqnarray}
\Lambda(\rho)=\sum_{\mu=1}^N K_\mu\rho K_\mu^\dagger,
\end{eqnarray}
where the Kraus operators $K_\mu$ satisfy not only $\sum_{\mu=1}^N K_\mu^\dagger K_\mu=\mathbb{I}$, but also $K_\mu\mathcal{I}K_\mu^\dagger\subset \mathcal{I}~\text{and}~K_\mu^\dag\mathcal{I}K_\mu\subset \mathcal{I}$ for every $K_\mu$ \cite{Winter, Yadin}. One sees by inspection that there is, at most, one nonzero element in each column and row of $K_\mu$, and $K_\mu$ are called strictly incoherent operators. From this, it is elementary to show that a projector is incoherent if it is of the form $\mathbb{P}_\mu=\sum_{i={\mu_0}}^{N_\mu}\ket{i}\bra{i}$ and we will denote $\mathbb{P}_\mu$ as a generic strictly incoherent projector. Hereafter, we will use $\Delta\rho=\sum_{i=1}^d\ket{i}\bra{i}\rho\ket{i}\bra{i}$ to denote the fully dephasing channel.

With the definition of strictly incoherent operations, we further introduce the notion of stochastic strictly incoherent operations \cite{Liu1}. A stochastic strictly incoherent operation is constructed by a subset of strictly incoherent Kraus operators. Without loss of generality, we denote the subset as $\{K_{1},K_{2},\dots, K_{L}\}$. Otherwise, we may renumber the subscripts of these Kraus operators. Then, a stochastic strictly incoherent operation, denoted as $\Lambda_s(\rho)$, is defined by
\begin{equation}
\Lambda_s(\rho)=\frac{\sum_{\mu=1}^L K_\mu\rho K_\mu^{\dagger}}{\Tr(\sum_{\mu=1}^LK_\mu\rho K_\mu^{\dagger})},
\label{lams}
\end{equation}
where $\{K_{1},K_{2},\dots, K_{L}\}$ satisfies $\sum_{\mu=1}^L K_\mu^{\dagger}K_\mu\leq \mathbb{I}$. Clearly, the state $\Lambda_s(\rho)$ is obtained with probability $P=\Tr(\sum_{\mu=1}^LK_\mu\rho K_\mu^{\dagger})$ under a stochastic strictly incoherent operation $\Lambda_s$, while state $\Lambda(\rho)$ is fully deterministic under a strictly incoherent operation $\Lambda$. Here, we emphasize that the stochastic transformation with $\sum_{\mu=1}^L K_\mu^{\dagger}K_\mu\leq \mathbb{I}$ means that a copy of $\Lambda_s(\rho)$ may be obtained from a copy of $\rho$ with probability $\text{P}=\Tr(\sum_{\mu=1}^LK_\mu\rho K_\mu^{\dagger})(\leq1)$. That is, the stochastic transformation runs the risk of failure with certain probability.

A functional $C$ can be taken as a coherence measure under strictly incoherent operations, if it satisfies the following four conditions \cite{Yadin}:
\begin{itemize}
\item[(C1)] Non-negativity: $C(\rho)\ge 0$, and $C(\rho)=0$ if and only if $\rho\in\mathcal{I}$;
\item[(C2a)] Monotonicity: $C$ does not increase under the action of strictly incoherent operations, i.e., $C(\rho)\ge C(\Lambda(\rho))$;
\item[(C2b)] Strong monotonicity: $C$ does not increase under selective strictly incoherent operations,
$C(\rho)\ge \sum_np_nC(\rho_n)$; where $p_n=\Tr(K_n\rho K_n^\dagger)$, $\rho_n=K_n\rho K_n^\dagger/p_n$;
\item[(C3)] Convexity: $C$ is a convex functional of the state, i.e., $\sum_nq_nC(\rho_n)\ge C(\sum_nq_n\rho_n)$ for any set of states $\{\rho_n\}$ and any probability distribution $\{q_n\}$.
\end{itemize}

\section{Coherence Filtration under Strictly Incoherent Operations} \label{III}

Let us move to consider the protocol of the coherence filtration, which can be formally presented as follows:~~
Given a state $\rho$, the aim of the protocol is to transform $\rho$ into some state $\rho^\prime$ by using some stochastic strictly incoherent operations $\Lambda_s$, which has the maximal fidelity with the maximally coherent state. In other words, we want to accomplish the transformation
\begin{eqnarray}
\rho\stackrel{\Lambda_s}{\longrightarrow}\rho^\prime
\end{eqnarray}
such that the value $F(\Lambda_s(\rho),\psi_d)$ with $\Lambda_s(\rho)=\rho^\prime$ is maximal, where the fidelity $F(\rho,\sigma)$ is defined as $F(\rho,\sigma):=\left(\Tr\sqrt{\rho^{1/2}\sigma\rho^{1/2}}\right)^2$ \cite{Uhlmann}. Hence, $F(\Lambda_s(\rho),\psi_d)=\Tr[\Lambda_s(\rho)\psi_d]$. With the above notions, we now present the following theorem.

\emph{Theorem} 1. Let $\Lambda_s$ be a stochastic strictly incoherent operation. Then
\begin{eqnarray}
\max_{\Lambda_s}\Tr[\Lambda_s(\rho)\psi_d]=\frac1d\uplambda_{\max}\left(\Delta\rho^{-\frac12}\rho\Delta\rho^{-\frac12}\right). \label{maximum}
\end{eqnarray}
Here, for a given state $\rho=\sum_{ij}\rho_{ij}\ket{i}\bra{j}$, the matrix $\Delta\rho=\sum_i\rho_{ii}\ket{i}\bra{i}$,
$(\Delta\rho)^{-\frac12}$ is the diagonal matrix with elements
$(\Delta\rho)^{-\frac12}_{ii}= \left\{
  \begin{array}{ll}
     \rho_{ii}^{-\frac12}, &\text{if} ~ \rho_{ii}\neq0;\\
    0,&\text{if}~ \rho_{ii}= 0.
  \end{array}\right.$, and $\uplambda_{\max}(A)$ denotes the maximal eigenvalue of $A$.

\emph{Proof}. First, we show that the maximum in Eq. (\ref{maximum}) is always achieved by a map $\Lambda_s$ with only one term,
\begin{eqnarray}
\Lambda_s^1(\rho)=\frac{K\rho K^\dag}{\Tr(K\rho K^\dag)},\label{single}
\end{eqnarray}
i.e., there is
\begin{eqnarray}
\max_{\Lambda_s}\Tr[\Lambda_s(\rho)\psi_d]=\max_{\Lambda_s^1}\Tr[\Lambda_s^1(\rho)\psi_d]. \label{simple}
\end{eqnarray}
To see this, let the form of $\Lambda_s(\rho)$ in Eq. (\ref{maximum}) be
\begin{eqnarray}
\Lambda_s(\rho)=\frac{\sum_{\mu=1}^L K_\mu\rho K_\mu^{\dagger}}{\Tr(\sum_{\mu=1}^LK_\mu\rho K_\mu^{\dagger})}. \label{stochastic}
\end{eqnarray}
By substituting Eq. (\ref{stochastic}) into Eq. (\ref{maximum}), one obtains that
\begin{eqnarray}
\Tr[\Lambda_s(\rho)\psi_d]=\frac{\sum_{\mu=1}^L\bra{\psi_d}K_\mu\rho K_\mu^{\dagger}\ket{\psi_d}}{\sum_{\mu=1}^L\Tr(K_\mu\rho K_\mu^{\dagger})}.
\end{eqnarray}
Next, let $p_\mu:=\bra{\psi_d}K_\mu\rho K_\mu^{\dagger}\ket{\psi_d}$ and $q_\mu:=\Tr(K_\mu\rho K_\mu^{\dagger})$. Then, given a finite pair of positive numbers, $(p_1,q_1)$, $(p_2,q_2)$, $\cdots$, $(p_L,q_L)$, one can see that
\begin{eqnarray}
\frac{\sum_{\mu=1}^Lp_\mu}{\sum_{\mu=1}^Lq_\mu}\leq\max_\mu\frac{p_\mu}{q_\mu}. \label{inequality}
\end{eqnarray}
To show this, let $\frac{p_\nu}{q_\nu}:=\max_\mu\frac{p_\mu}{q_\mu}$. It is direct to obtain the result in Eq. (\ref{inequality}) from the following calculations
\begin{eqnarray}
\frac{p_\nu}{q_\nu}\sum_{\mu=1}^Lq_\mu-\sum_{\mu=1}^Lp_\mu\geq\sum_{\mu=1}^L\left(\frac{p_\mu}{q_\mu}q_\mu-p_\mu\right)=0.
\end{eqnarray}
Thus, the maximum in Eq. (\ref{maximum}) can always be obtained by only considering $\Lambda_s^1(\rho)$, which implies the relation in Eq. (\ref{simple}).

Second, we show that
\begin{eqnarray}
\max_{\Lambda_s^1}\Tr[\Lambda_s^1(\rho)\psi_d]\leq\frac1d\uplambda_{\max}\left(\Delta\rho^{-\frac12}\rho\Delta\rho^{-\frac12}\right). \label{step2}
\end{eqnarray}
To see this, let us consider the structure of a strictly incoherent operator $K$. From the definition of the strictly incoherent operator, we obtain that any strictly incoherent operator $K$ can always be decomposed into $K=P_\pi K^\prime$,
where $P_\pi$ is a permutation matrix and $K^\prime=\text{diag}(a_1,a_2...,a_d)$ is a diagonal matrix. On the other hand, it is direct to obtain that for any permutation matrix $P_\pi$, there is $P_\pi\ket{\psi_d}=\ket{\psi_d}$. With these results, by direct calculations, we obtain
\begin{eqnarray}
\Tr[\Lambda_s^1(\rho)\psi_d]=\frac{\sum_{i,j=1}^da_i\rho_{ij}a^*_j}{d\sum_{j=1}^d\abs{a_j}^2\rho_{jj}}. \label{sum}
\end{eqnarray}
Let us further introduce a normalized vector,
\begin{eqnarray}
\ket{\varphi}:=\frac1{\sqrt{\sum_j\abs{a_j}^2\rho_{jj}}}(\Delta\rho)^\frac12(a_1^*,a_2^*,\cdots,a_d^*)^t, \label{vector}
\end{eqnarray}
where $(\cdot)^t$ is the transpose. By substituting Eq. (\ref{vector}) into Eq. (\ref{sum}), we obtain
\begin{eqnarray}
\Tr[\Lambda_s^1(\rho)\psi_d]=\frac{\bra{\varphi}\Delta\rho^{-\frac12}\rho\Delta\rho^{-\frac12}\ket{\varphi}}d.
\end{eqnarray}
Since $\bra{\varphi}\Delta\rho^{-\frac12}\rho\Delta\rho^{-\frac12}\ket{\varphi}\leq\uplambda_{\max}(\Delta\rho^{-\frac12}\rho\Delta\rho^{-\frac12})$ with $\uplambda_{\max}(A)$ being the maximal eigenvalue of $A$, we obtain that
\begin{eqnarray}\label{inequality2}
\max_{\Lambda_s^1}\Tr[\Lambda_s^1(\rho)\psi_d]\leq\frac1d\uplambda_{\max}\left(\Delta\rho^{-\frac12}\rho\Delta\rho^{-\frac12}\right),
\end{eqnarray}
which is the relation in Eq. (\ref{step2}).

Third, we show that the upper bound in Eq. (\ref{inequality2}) can be achieved by some $\Lambda_s^1$, i.e., there is
\begin{eqnarray}
\max_{\Lambda_s^1}\Tr[\Lambda_s^1(\rho)\psi_d]=\frac1d\uplambda_{\max}\left(\Delta\rho^{-\frac12}\rho\Delta\rho^{-\frac12}\right). \label{step3}
\end{eqnarray}
To show this, let us denote by
\begin{eqnarray}
\ket{\uplambda_{\max}}:=\frac1{\sum_{j=1}^d\abs{c_j}^2}(c_1,c_2,\cdots,c_d)^t
\end{eqnarray}
a normalized eigenvector corresponding to the largest eigenvalue of $\Delta\rho^{-\frac12}\rho\Delta\rho^{-\frac12}$, i.e., there is
\begin{eqnarray}
\Delta\rho^{-\frac12}\rho\Delta\rho^{-\frac12}\ket{\uplambda_{\max}}=
\uplambda_{\max}(\Delta\rho^{-\frac12}\rho\Delta\rho^{-\frac12})\ket{\uplambda_{\max}}.
\end{eqnarray}
Let us define $a_j:=\frac{c_j^*}{\sqrt{\rho_{jj}}}$ if $\rho_{jj}\neq0$ and $a_j:=0$ if $\rho_{jj}=0$. Then, let us choose the strictly incoherent operator as
\begin{eqnarray}
K=\text{diag}(a_1,a_2,\cdots,a_d).\label{Kraus}
\end{eqnarray}
By direct calculations, we immediately obtain
\begin{eqnarray}
\Tr[\Lambda_s^1(\rho)\psi_d]=\frac1d\uplambda_{\max}\left(\Delta\rho^{-\frac12}\rho\Delta\rho^{-\frac12}\right),
\end{eqnarray}
which is the relation in Eq. (\ref{step3}). This completes the proof of the theorem. ~~~~~~~~~~~~~~~~~~~~~~~~~~~~~~~~~~~~~~~~~~~~~~~~~~~~~~~~~~~~~~~~~~~~~~~~~~~~~~~~~~~~~~~$\blacksquare$

The theorem mentioned above, along with its proof, addresses the aforementioned question above: How can we convert a given state $\rho$ into a state $\rho^\prime$ that has the highest possible fidelity to the maximally coherent state by using stochastic strictly incoherent operations denoted as $\Lambda_s$? The maximum fidelity achievable with the maximally coherent state is determined by Eq. (\ref{maximum}). The required operations can be chosen based on Eq. (\ref{single}) and the corresponding Kraus operator can be selected as described in Eq. (\ref{Kraus}).

\section{relationships with coherence measures} \label{IV}

In this section, we will present several results relating to the above theorem.

The first is that we can provide an operational interpretation of the $\Delta$ robustness of coherence \cite{Chitambar2}, which is defined as
\begin{eqnarray}
R(\rho\|\Delta\rho):=\text{min}\{\uplambda|\rho\leq\uplambda\Delta\rho\}.
\end{eqnarray}
 This leads to the following theorem.

\emph{Theorem} 2. For a given $d$-dimensional density matrix $\rho$, the maximum achievable fidelity in Eq. (\ref{maximum}) is determined by the $\Delta$ robustness of coherence. In other words, the expression can be written as follows:
\begin{eqnarray}
\max_{\Lambda_s}\Tr[\Lambda_s(\rho)\psi_d]=\frac1d R(\rho\parallel\Delta\rho).\label{corollary1}
\end{eqnarray}

\emph{Proof}. To see this, we first show that $\text{supp}(\rho)\subset\text{supp}(\Delta\rho)$, where $\text{supp}(A)$ means the support of $A$ \cite{notesupp}. Let the spectral decomposition of $\rho$ be $\rho=\sum_{j=1}^l\uplambda_j\ket{\uplambda_j}\bra{\uplambda_j}$, where $\uplambda_j>0$. By the definition of support, we obtain that $\text{supp}(\rho)=\text{span}\{\ket{\uplambda_1},\ket{\uplambda_2},\cdots,\ket{\uplambda_l}\}$ and $\text{supp}(\Delta\rho)=\text{span}\{\ket{1},\ket{2},\cdots,\ket{n}\}$ with $n\leq d$. With these notions, we prove $\text{supp}(\rho)\subset\text{supp}(\Delta\rho)$ by contradiction. Suppose there exists some $\ket{\uplambda_j}\notin\text{supp}(\Delta\rho)$, then there is some $\ket{m}\notin\text{supp}(\Delta\rho)$ while $\langle m\ket{\uplambda_j}\neq0$ for some $j$. We further obtain that $\bra{m}\rho\ket{m}\geq\uplambda_j\abs{\langle m\ket{\uplambda_j}}^2>0$. But this is in contradiction with $\ket{m}\notin\text{supp}(\Delta\rho)$. Thus, we obtain that
\begin{eqnarray}
\text{supp}(\rho)\subset\text{supp}(\Delta\rho).
\end{eqnarray}

Next, with $\text{supp}(\rho)\subset\text{supp}(\Delta\rho)$, one can see that $\rho\leq\uplambda\Delta\rho$ is equivalent to \begin{eqnarray}
\Delta\rho^{-\frac12}\rho\Delta\rho^{-\frac12}\leq\uplambda\mathbb{I}. \label{bound}
\end{eqnarray}
Further, it is direct to see that the smallest $\uplambda$ achieving the inequality in Eq. (\ref{bound}) is $\uplambda_{\max}\left(\Delta\rho^{-\frac12}\rho\Delta\rho^{-\frac12}\right)$. By using Theorem 1, we obtain the result in Eq. (\ref{corollary1}). ~~~~~~~~~~~~~~~~~~~~~~~~~~~~~~~~~~~~~~~~~~~~~~~~~$\blacksquare$

We would like to stress that we adopt the term ``operational interpretation" by following Refs. \cite{Winter,Chitambar,Napoli}. This term describes a scenario in which a particular quantity or resource measure provides a quantitative depiction of a fundamental parameter within an operational task or a task that benefits from the resource. In the Theorem 2, we can present an operational interpretation of the $\Delta$ robustness of coherence. This interpretation demonstrates that the quantification of filtration coherence is precisely captured by the $\Delta$ robustness of coherence.

The second is that $\max_{\Lambda_s}\Tr[\Lambda_s(\rho)\psi_d]$ obtains its minimum if and only if $\rho$ is an incoherent state and $\max_{\Lambda_s}\Tr[\Lambda_s(\rho)\psi_d]$ obtains its maximum if and only if $\rho$ is a pure coherent state with its coherence rank being $d$. Here, for a pure state $\varphi$ (not necessary normalized), the coherence rank of it $C_r(\varphi)$ is the rank of $\Delta\varphi$ \cite{Winter,note}, i.e.,
\begin{eqnarray}
C_r(\varphi):=\text{Rank}(\Delta\varphi).
\end{eqnarray}
This arrives at the following theorem.

\emph{Theorem} 3. Let $\rho$ be an arbitrary $d$-dimensional density matrix. Then there is
\begin{eqnarray}
\frac1d\leq\max_{\Lambda_s}\Tr[\Lambda_s(\rho)\psi_d]\leq1, \label{corollary2}
\end{eqnarray}
where $\max_{\Lambda_s}\Tr[\Lambda_s(\rho)\psi_d]=\frac1d$ if and only if $\rho$ is an incoherent state and $\max_{\Lambda_s}\Tr[\Lambda_s(\rho)\psi_d]=1$ if and only if $\rho$ is a pure coherent state with its coherence rank being $d$.

\emph{Proof}. First, we show that $\frac1d\leq\max_{\Lambda_s}\Tr[\Lambda_s(\rho)\psi_d]$. For the sake of simplicity, we denote
\begin{eqnarray}
C_s(\rho):=\max_{\Lambda_s}\Tr[\Lambda_s(\rho)\psi_d].
\end{eqnarray}
Let $\Lambda_s^\prime$ and $\Lambda_s$ be two stochastic strictly incoherent operations. Then, by the definition of stochastic strictly incoherent operations, it is direct to examine that $\Lambda_s^\prime\circ\Lambda_s$ is also a stochastic strictly incoherent operation. Then, there is
\begin{eqnarray}
C_s\left(\Lambda_s(\rho)\right)\leq C_s(\rho). \label{monotone}
\end{eqnarray}
By using the facts that (i) the set of strictly incoherent operations is a subset of stochastic strictly incoherent operations, (ii) any incoherent state can be obtained from a coherent state by using strictly incoherent operations, and (iii) any incoherent state can be obtained from another incoherent state by using strictly incoherent operations, one can see that the minimum of $C_s(\rho)$ is obtained when $\rho$ is an incoherent state. Thus, the minimum of $C_s(\rho)$ can be obtained when $\Lambda_s(\rho)=\ket{1}\bra{1}$. Therefore, there is $\frac1d\leq C_s(\rho)$, i.e., $\frac1d\leq\max_{\Lambda_s}\Tr[\Lambda_s(\rho)\psi_d]$. Then, we show that there is $\frac1d<\max_{\Lambda_s}\Tr[\Lambda_s(\rho)\psi_d]$ when $\rho$ is a coherent state. To see this, suppose $\rho$ is a coherent state, then there is some $\rho_{ij}\neq0$. Let $\rho^\prime:=\Lambda_s(\rho)=\frac{K\rho K^\dag}{\Tr(K\rho K^\dag)}$, where $K=\ket{i}\bra{i}+\ket{j}\bra{j}$. By using Eq. (\ref{monotone}), we obtain that $C_s(\rho^\prime)\leq C_s(\rho)$, with
\begin{eqnarray}
\rho^\prime=\frac1{\rho_{ii}+\rho_{jj}}\begin{pmatrix}
    \rho_{ii}&\rho_{ij}\\
    \rho_{ji}&\rho_{jj}
  \end{pmatrix}.
\end{eqnarray}
We next show that $C_s(\rho^\prime)>\frac1d$. To this end, one can see, from Theorem 1, that this is equivalent to showing that $\uplambda_{\max}\left(\Delta{\rho^\prime}^{-\frac12}{\rho^\prime}\Delta{\rho^\prime}^{-\frac12}\right)>1$. One can be obtained by calculating the eigenvalues of $\Delta{\rho^\prime}^{-\frac12}{\rho^\prime}\Delta{\rho^\prime}^{-\frac12}$.
This means that $\frac1d\leq\max_{\Lambda_s}\Tr[\Lambda_s(\rho)\psi_d]$ and $\max_{\Lambda_s}\Tr[\Lambda_s(\rho)\psi_d]=\frac1d$ if and only if $\rho$ is an incoherent state.

Next, we show that $\max_{\Lambda_s}\Tr[\Lambda_s(\rho)\psi_d]\leq1$. To see this, we consider the properties of the matrix $\mathbb{B}_\rho:=\Delta\rho^{-\frac12}\rho\Delta\rho^{-\frac12}$. It is direct to see that $\mathbb{B}_\rho$ is a positive semi-definite matrix and all its non-zero diagonal elements are 1. Let $\uplambda_j$ with $1\leq j\leq d$ be the nonzero eigenvalues of $\mathbb{B}_\rho$. Then there is $\sum_j\uplambda_j=d$. Thus, one can see that $\uplambda_{\max}(\mathbb{B}_\rho)\leq d$. Since $\uplambda_{\max}(\mathbb{B}_\rho)=d$ if and only if the rank of $\mathbb{B}_\rho$ is one and $\text{supp}(\rho)\subset\text{supp}(\Delta\rho)$, we immediately obtain that the rank of $\rho$ is one and its coherence rank is $d$. This completes the proof of the theorem. ~~~~~~~~~~~~~~~~$\blacksquare$

With the above results, we find that we can obtain a coherence measure from the task of coherence filtration, which is defined as
\begin{eqnarray}
C_m(\rho):=\max_{\Lambda_s}\Tr[\Lambda_s(\rho)\psi_d]-\frac1d.
\end{eqnarray}
This leads to the following theorem.

\emph{Theorem} 4. The functional $C_m(\rho)$ is a coherence measure satisfying the conditions (C1)-(C3).

\emph{Proof}. First, we show that $C_m(\rho)$ satisfies the condition (C1). By theorem 3, which says that $\max_{\Lambda_s}\Tr[\Lambda_s(\rho)\psi_d]\geq\frac1d$, where the equality is obtained if and only if $\rho$ is an incoherent state, we immediately obtain that $C_m(\rho)\geq0$ and
\begin{eqnarray}
C_m(\rho)=0 ~~\text{if~and~only~if}~~\rho\in\mathcal{I}.
\end{eqnarray}

Second, we show that $C_m(\rho)$ satisfies the condition (C2b). Let $\{\Lambda_s^k\}$ be a set of stochastic strictly incoherent operations, whose sum $\sum_kP_k\Lambda_s^k(\rho)=:\Lambda(\rho)$ defines a (trace-preserving) strictly incoherent operation. Then, by Eq. (\ref{monotone}), we obtain that $\sum_kP_kC_s(\Lambda_s^k(\rho))\leq C_s(\rho)$, which further implies the condition (C2b), i.e., \begin{eqnarray}
\sum_kP_kC_m(\Lambda_s^k(\rho))\leq C_m(\rho).
\end{eqnarray}

Third, we show that $C_m(\rho)$ satisfies the condition (C3). Let $\rho_1$ and $\rho_2$ be two states and $\rho=p\rho_1+(1-p)\rho_2$ with $0\leq p\leq1$. For the state $\rho$, let $\Gamma_s$ be a stochastic strictly incoherent operation achieving the maximum in Eq. (\ref{maximum}), i.e., $\max_{\Lambda_s}\Tr[\Lambda_s(\rho)\psi_d]=\Tr[\Gamma_s(\rho)\psi_d]$. Then, one can see that
\begin{eqnarray}
&&\max_{\Lambda_s}\Tr[\Lambda_s(\rho)\psi_d]=\Tr[\Gamma_s(\rho)\psi_d]\nonumber\\
&&=\Tr\left(\Gamma_s(p\rho_1+(1-p)\rho_2)\psi_d\right)\nonumber\\
&&=p\Tr\left(\Gamma_s(\rho_1)\psi_d\right)+(1-p)\Tr\left(\Gamma_s(\rho_2)\psi_d\right)\nonumber\\
&&\leq p\max_{\Lambda_s}\Tr[\Lambda_s(\rho_1)\psi_d]+(1-p)\max_{\Lambda_s}\Tr[\Lambda_s(\rho_2)\psi_d].\nonumber\\
\end{eqnarray}
This implies the condition (C3),
\begin{eqnarray}
C_m(\rho)\leq pC_m(\rho_1)+(1-p)C_m(\rho_2).
\end{eqnarray}

Since conditions (C2b) and (C3) imply the condition (C2a), we obtain that $C_m(\rho)$ is a coherence measure. This completes the proof of the theorem. ~~~~~~~~~~~~~~~~~~~~~~~~~~~~~~~~~~~~~~~~~~~~~~~~~~~~~~~$\blacksquare$

We would like to point out that the quantity $R(\rho\|\Delta\rho)$ can be viewed as a coherence monotone which extends
the coherence rank to mixed states. Here, we say a functional is a coherence monotone if it satisfies the conditions (C2a), (C2b), and (C3).

Finally, we discuss the relationship between $C_m(\rho)$ and the state transformations under stochastic strictly incoherent operations. This leads to the following corollary.

\emph{Corollary} 1. There is a stochastic strictly incoherent operation $\Lambda_s$ such that $\Lambda_s(\varphi_1)=\varphi_2$ if and only if
$C_m(\varphi_1)\geq C_m(\varphi_2)$. However, for mixed states, this condition is only a necessary one.

\emph{Proof}. By using a result in Ref. \cite{Winter} which says that the transformation is impossible when the coherence rank of the target state is larger than that of the initial state, but otherwise the conversion is achievable by using some stochastic strictly incoherent operation, we then obtain that if $\Lambda_s(\varphi_1)=\varphi_2$, there is $C_r(\varphi_1)\geq C_r(\varphi_2)$. This further implies $C_m(\varphi_1)\geq C_m(\varphi_2)$.

For the mixed states case, the necessary part of this condition can be obtained from the expression in Eq. (\ref{monotone}). Next, we show that this condition is not a sufficient one. To see this, let us consider the following two three-dimensional states:
\begin{eqnarray}\label{state}
  \rho_1=\frac1{15}\begin{pmatrix}
    5&4&4\\
    4&5&4\\
   4&4&5
  \end{pmatrix}
  \end{eqnarray}
  and
  \begin{eqnarray}
  \rho_2=\frac12\begin{pmatrix}
    1&1&0\\
    1&1&0\\
   0&0&0
  \end{pmatrix}.
\end{eqnarray}
By using Theorem 1 and direct calculations, we obtain that $C_m(\rho_1)=0.533>C_m(\rho_2)=0.333$. On the other hand, let us recall a result in Ref. \cite{Liu3}, which says that a pure coherent state $\ket{\varphi}$ can be obtained from a mixed state $\rho$ by using stochastic strictly incoherent operations if and only if there is an incoherent projector $\mathbb{P}$ with  the coherence rank of $\mathbb{P}\rho\mathbb{P}$ being greater than or equal to that of $\ket{\varphi}$. However, it is direct to examine that there is no incoherent projector $\mathbb{P}$ such that the coherence rank of $\mathbb{P}\rho_1\mathbb{P}$ is greater than or equal to 2. Thus, we cannot transform $\rho_1$ into $\rho_2$ via stochastic strictly incoherent operations. This completes the proof of the corollary.~~~~~~~~~~~ $\blacksquare$

\section{conclusions} \label{V}

 To summarize, we have studied the task of coherence filtration in this paper. Specifically, the aim of this task is to transform a given state $\rho$ into another one $\rho^\prime$ whose fidelity with the maximally coherent state is maximal by using stochastic strictly incoherent operations. Interestingly, we have found that this maximal fidelity between $\rho^\prime$ and the maximally coherent state is given by the $\Delta$ robustness of coherence $R(\rho\|\Delta\rho)$ in Theorems 1 and 2. Thus, we provide the $\Delta$ robustness of coherence an operational interpretation. Furthermore, we obtain a coherence measure from the task of coherence filtration in Theorems 3 and 4. Finally, we discuss the relation between $C_m(\rho)$ and the state transformations under stochastic strictly incoherent operations in Corollary 1.

 In passing, we would like to point out that strictly incoherent Kraus operators can always be constructed by the system interacting with an ancilla and a general experimental setting has been suggested based on an interferometer in Ref. \cite{Yadin} and an experimental implementation of it has  recently been presented in Ref. \cite{Xiong}. Thus, our scheme of coherence filtration can be experimentally demonstrated by using the setup in \cite{Xiong}.

\section*{Acknowlegements}

This work is supported by the NSFC (Grant No. 12088101), and NSAF (Grant No. U1930403). C.L.L. acknowledges support from the China Postdoctoral Science Foundation, Grant No. 2021M690324.


\begin{thebibliography}{99}
\bibitem{Nielsen} M. A. Nielsen and I. L. Chuang, Quantum Computation and Quantum Information, (Canbrudge University Press, Cambridge, 2000).
\bibitem{Shor} P. W. Shor, \text{SIAM J. Comput.} \textbf{26}, 1484 (1997).
\bibitem{Grover} L. K. Grover, \text{Phys. Rev. Lett.} \textbf{79}, 325 (1997).
\bibitem{Bennett} C. H. Bennett and G. Brassard, \text{Theor. Comput. Sci.} \textbf{560}, 7 (2014).
\bibitem{Giovannetti} V. Giovannetti, S. Lloyd, and L. Maccone,  \text{Science} \textbf{306}, 1330 (2004).
\bibitem{Giovannetti1} V. Giovannetti, S. Lloyd, and L. Maccone,  \text{Nat. Photonics} \textbf{5}, 222 (2011).
\bibitem{Lambert} N. Lambert, Y.-N. Chen, Y.-C. Cheng, C.-M. Li,  G.-Y. Chen, and F. Nori, \text{Nat. Phys.}  \textbf{9}, 10 (2013).
\bibitem{Aberg1} J. {\AA}berg, arXiv:quant-ph/0612146.
\bibitem{Baumgratz} T. Baumgratz, M. Cramer, and M. B. Plenio, \text{Phys. Rev. Lett.} \textbf{113}, 140401 (2014).
\bibitem{Streltsov} A. Streltsov, G. Adesso, and M. B. Plenio, \text{Rev. Mod. Phys.} \textbf{89}, 041003 (2017).
\bibitem{Fan} M.-L. Hu, X. Hu, J.-C. Wang, Y. Peng, Y.-R. Zhang, and H. Fan, \text{Phys. Rep.} \textbf{762-764}, 1 (2018).
\bibitem{Wu} K.-D. Wu, A. Streltsov, B. Regula, G.-Y. Xiang, C.-F. Li, and G.-C. Guo, Adv. Quantum Technol. \textbf{4}, 2100040 (2021)
\bibitem{Horodecki} M. Horodecki and J. Oppenheim, \text{Int. J. Mod. Phys. B} \textbf{27}, 1345019 (2013).
\bibitem{Brandao} F. G. S. L. Brand\~{a}o and G. Gour, \text{Phys. Rev. Lett.} \textbf{115}, 070503 (2015).
\bibitem{Chitambar} E. Chitambar and G. Gour, \text{Rev. Mod. Phys.} \textbf{91}, 025001 (2019).
\bibitem{Levi} F. Levi and F. Mintert, \text{New J. Phys.} \textbf{16}, 033007 (2014).
\bibitem{Winter} A. Winter and D. Yang, \text{Phys. Rev. Lett.} \textbf{116}, 120404 (2016).
\bibitem{Yadin} B. Yadin, J. Ma, D. Girolami, M. Gu, and V. Vedral, \text{Phys. Rev. X} \textbf{6}, 041028 (2016).
\bibitem{Chitambar1} E. Chitambar and G. Gour, \text{Phys. Rev. Lett.} \textbf{117}, 030401 (2016).
\bibitem{Chitambar2} E. Chitambar and G. Gour, \text{Phys. Rev. A} \textbf{94}, 052336 (2016).
\bibitem{Marvian} I. Marvian and R. W. Spekkens, \text{Phys. Rev. A} \textbf{94}, 052324 (2016).
\bibitem{Vicente} J. I. de Vicente and A. Streltsov, \text{J. Phys. A: Math. Theor.} \textbf{50}, 045301 (2017).
\bibitem{Lami1} L. Lami, B. Regula, and G. Adesso, \text{Phys. Rev. Lett.} \textbf{122}, 150402 (2019).
\bibitem{Zhao} Q. Zhao, Y. Liu, X. Yuan, E. Chitambar, and A. Winter, \text{IEEE Trans. Inf. Theory} \textbf{65}, 6441  (2019).
\bibitem{Lami} L. Lami,  \text{IEEE Trans. Inf. Theory} \textbf{66}, 2165 (2020).
\bibitem{Bu}K. F. Bu, U. Singh,  S.-M. Fei,  A. K. Pati, and J. D. Wu, \text{Phys. Rev. Lett.} \textbf{119}, 150405 (2017).
\bibitem{Liu1} C. L. Liu, Y.-Q. Guo, and D. M. Tong, \text{Phys. Rev. A} \textbf{96}, 062325 (2017).
\bibitem{Regula} B. Regula, L. Lami, and A. Streltsov, \text{Phys. Rev. A} \textbf{98}, 052329 (2018).
\bibitem{Regula1} B. Regula, K. Fang, X. Wang, and G. Adesso, \text{Phys. Rev. Lett.} \textbf{121}, 010401 (2018).
\bibitem{Fang} K. Fang, X. Wang, L. Lami, B. Regula, and G. Adesso, \text{Phys. Rev. Lett.} \textbf{121}, 070404 (2018).
\bibitem{Torun} G. Torun, L. Lami, G. Adesso, and A. Yildiz, \text{Phys. Rev. A} \textbf{99}, 012321 (2019).
\bibitem{Liu2} C. L. Liu and D. L. Zhou,  \text{Phys. Rev. Lett.} \textbf{123}, 070402 (2019).
\bibitem{Chen} S. Chen, X. Zhang, Y. Zhou, and Q. Zhao, \text{Phys. Rev. A} \textbf{100}, 042323 (2019)
\bibitem{Liu3}C. L. Liu and D. L. Zhou,  \text{Phys. Rev. A} \textbf{101}, 012313 (2020).
\bibitem{Zhang}S. Zhang, Y. Luo, L.-H. Shao, Z. Xi, and H. Fan, \text{Phys. Rev. A} \textbf{102}, 052405 (2020).
\bibitem{Liu4} C. L. Liu and C. P. Sun,  \text{Phys. Rev. Research} \textbf{3}, 043220 (2021).
\bibitem{Liu5} C. L. Liu and C. P. Sun,  \text{Phys. Rev. Research} \textbf{4}, 023199 (2022).
\bibitem{Dur} W. D\"{u}r and H. J. Briegel, \text{Rep. Prog. Phys.} \textbf{70}, 1381 (2007).
\bibitem{Uhlmann} A. Uhlmann, Rep. Math. Phys. \textbf{9}, 273 (1976).
\bibitem{notesupp} Here, for an operator A, the support of it is the vector space spanned by the eigenvectors of A with non-zero eigenvalue \cite{Nielsen}.
\bibitem{Napoli} C. Napoli, T. R. Bromley, M. Cianciaruso, M. Piani, N. Johnston, and G. Adesso, \text{Phys. Rev. Lett.} \textbf{116}, 150502 (2016).
\bibitem{note} We should note that the coherence rank, as defined in \cite{Winter}, denoted as $C_r(\varphi):=R$ with $\ket{\varphi}=\sum_{j=1}^Rc_j\ket{j}$ and $c_j\neq0$ for all $j=1,\cdots,R$, does not satisfy (C1).
\bibitem{Xiong} S.-J. Xiong, Z. Sun, Q.-P. Su, Z.-J. Xi, L. Yu, J.-S. Jin, J.-M. Liu, F. Nori, and C.-P. Yang,  \text{Optica} \textbf{8}, 1003(2021).
\end{thebibliography}
\end{document}